\pdfoutput=1
\documentclass[conference,10pt,letterpaper]{IEEEtran}
\IEEEoverridecommandlockouts
\usepackage{cite}
\usepackage{amsmath,amsfonts,bm,amssymb,pifont}
\usepackage{mathtools}
\usepackage{comment}
\usepackage{graphicx}
\usepackage{xcolor}
\usepackage[caption=false,font=footnotesize]{subfig}
\usepackage{booktabs}
\usepackage{array}
\usepackage{multirow}
\usepackage{enumitem}
\usepackage{placeins}
\usepackage{algorithm}
\usepackage{algpseudocode}

\setlist[itemize,1]{itemsep=0pt,partopsep=0pt,parsep=\parskip,topsep=2pt}

\providecommand{\Description}[1]{}
\usepackage[hidelinks]{hyperref}
\hypersetup{
  pdftitle={GRADPlace: Geometry-Based Routing-Augmented Differentiable Placement for Routability Optimization},
  pdfauthor={Xurong Shen, Xueyan Zhao, Zhipeng Huang, Chaoyu Xing, Biwei Xie},
  pdfsubject={},
  pdfkeywords={routability-driven placement, analytical placement, differentiable optimization, VLSI physical design}
}

\begin{document}

\title{GRADPlace: Geometry-Based Routing-Augmented Differentiable Placement for Routability Optimization}
\author{%
\IEEEauthorblockN{%
  Xurong Shen$^{1,2,\ast}$,
  Xueyan Zhao$^{2,\ast}$,
  Zhipeng Huang$^{3}$,
  Chaoyu Xing$^{1,2}$, and
  Biwei Xie$^{2,\dagger}$%
}
\IEEEauthorblockA{$^{1}$ShanghaiTech University, Shanghai, China\\
$^{2}$Institute of Computing Technology, Chinese Academy of Sciences, Beijing, China\\
$^{3}$Beijing Institute of Open Source Chip, Beijing, China\\
Emails: shawn@shanghaitech.edu.cn; zhaoxueyan21b@ict.ac.cn; huangzhipeng@bosc.ac.cn;\\
xingchy2025@shanghaitech.edu.cn; xiebiwei@ict.ac.cn}
\IEEEauthorblockA{$^{\ast}$These authors contributed equally to this work.\quad $^{\dagger}$Corresponding author.}
}

\maketitle

\begin{abstract}
Routability-driven placement needs gradients that show how cell movement redistributes routing demand. Cell inflation exploits routing feedback only indirectly, and differentiable methods rely on black-box congestion predictors or on virtual-cell congestion fields frozen between routing updates. We present GRADPlace, which represents probability-weighted candidate routes as movable horizontal and vertical segments. Routing feedback sets their probabilities, while their geometry and expected demand evolve with cell coordinates. An electrostatic routing objective propagates gradients through this geometry to connected cells, complementing cell inflation. On the ISPD 2015 benchmarks, our reproduced Xplace-route and published DCGP results have on average 3.37$\times$ and 2.56$\times$ as many design-rule violations as GRADPlace, with more vias and comparable routed wirelength.

\end{abstract}

\begin{IEEEkeywords}
routability-driven placement, analytical placement, differentiable optimization, VLSI physical design
\end{IEEEkeywords}

\section{Introduction}
\label{sec:Introduction}

Placement strongly influences the routing demand that must be accommodated by limited interconnect resources. Although modern analytical placers efficiently optimize wirelength and cell density, a placement with short wirelength and low cell overlap can still suffer from routing congestion and design-rule violations (DRVs) after detailed routing~\cite{lu2014eplace,cheng2019replace,lin2021dreamplace,liu2024xplace}. Routability-driven placement therefore requires optimization signals that connect cell movement to the spatial distribution of routing demand~\cite{chow2016placement,huang2018ntuplace4dr}.

\suppressfloats[t]
\begin{figure}[t]
    \centering
    \subfloat[Virtual-cell congestion gradient in DCGP~\cite{li2025dcgp}.\label{fig:dcgp_representative_point}]{%
        \includegraphics[width=0.48\columnwidth]{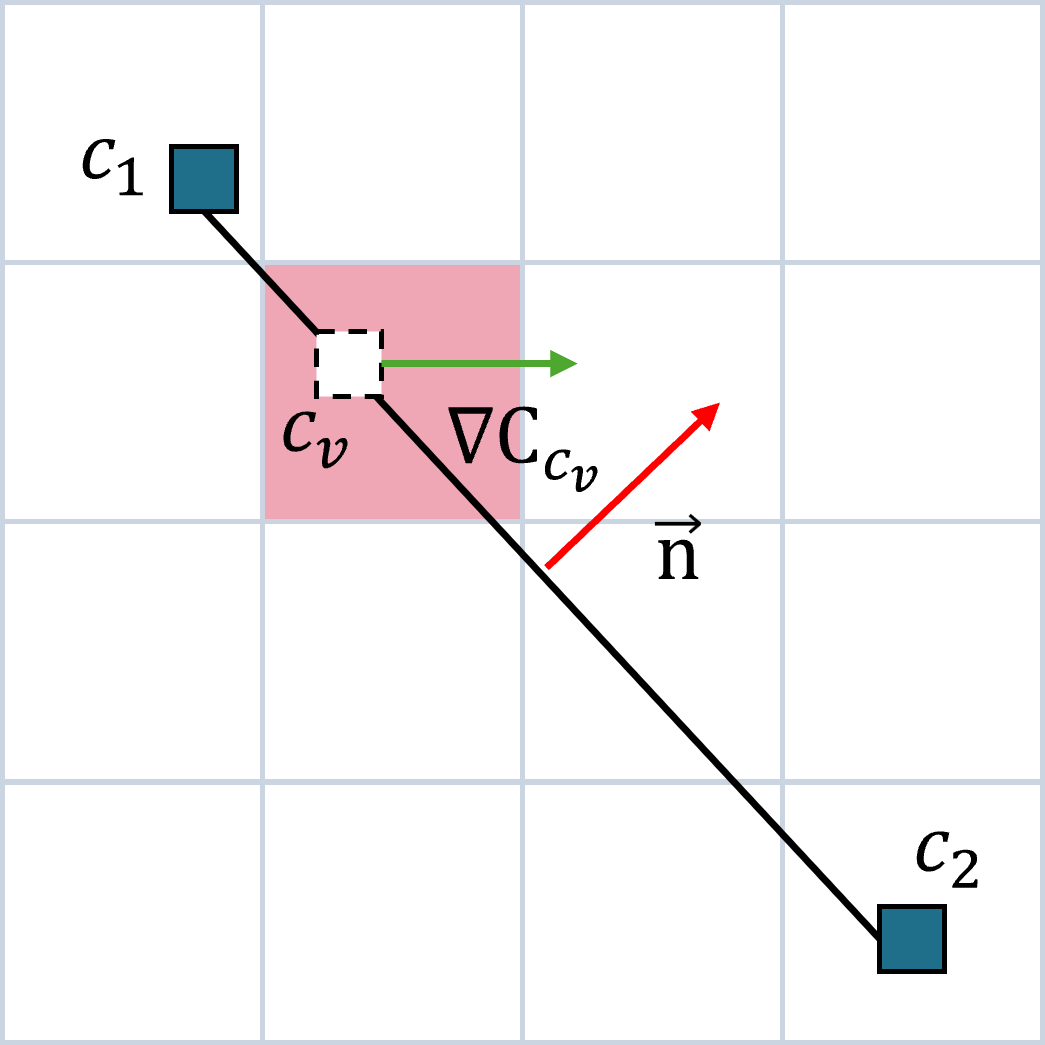}}
    \hfill
    \subfloat[Directional-segment congestion gradient in GRADPlace.\label{fig:gradplace_segments}]{%
        \includegraphics[width=0.48\columnwidth]{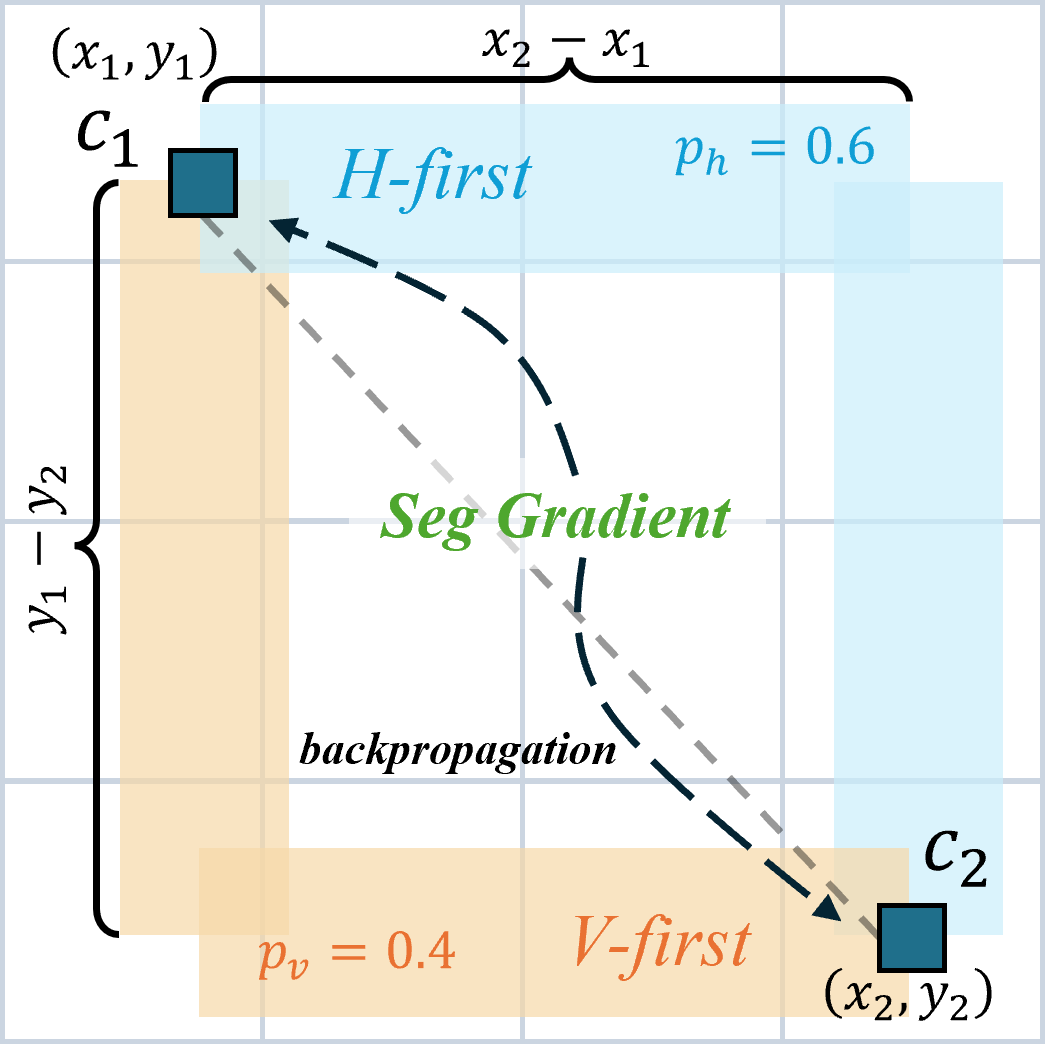}}
    \caption{Congestion-gradient models for a two-pin net. (a) DCGP samples $\nabla C_{c_v}$ at a virtual cell $c_v$, projects it onto the net normal $\vec{\mathbf n}$, and distributes it to endpoints by distance. (b) GRADPlace integrates directional fields over probability-weighted H-first and V-first segments and backpropagates through segment geometry.}
    \Description{Two grid-based routing illustrations. The first shows a congestion gradient sampled at a representative point on a two-pin net. The second shows directional segments induced by horizontal-first and vertical-first candidates, with their congestion gradients propagated to the connected cells.}
    \label{fig:representative_vs_segment}
\end{figure}

Recent work has shown that local cell crowding and global routing-demand concentration require complementary mechanisms~\cite{chen2025ruplace,li2025dcgp}. Classical probabilistic congestion estimators spread net demand over a bounding box, either uniformly as in RUDY~\cite{spindler2007rudy} or by weighting all shortest routes equally as in Lou et al.~\cite{lou2002probabilistic}, and cell-inflation methods use such estimates or global-routing maps to create whitespace in congested regions~\cite{he2013ripple,hsu2014ntuplace4h,lin2022ppom,cai2023puffer}. RUPlace~\cite{chen2025ruplace} combines modularity-based cell inflation with refinement guided by routed wires, alternating routing feedback and incremental placement. However, a region may be congested by pass-through nets whose pins lie elsewhere, so local density adjustment provides only indirect control over their demand.

Other approaches derive cell-movement guidance from differentiable models~\cite{liu2021explicitroute,hou2024routeplacer,lu2026c3po}. Liu et al.~\cite{liu2021explicitroute} and RoutePlacer~\cite{hou2024routeplacer} differentiate learned neural or GNN congestion predictors with respect to cell coordinates, but the learned predictors carry no explicit routing geometry. C3PO~\cite{lu2026c3po} concurrently optimizes timing, routability, and wirelength.

DCGP~\cite{li2025dcgp} places virtual cells on two-pin nets and uses an electrostatic field built from external routing feedback to guide connected cells. However, its congestion map remains frozen between feedback updates and does not reflect demand redistribution as cells move. Moreover, virtual-cell gradients are projected and distance-weighted to endpoints rather than propagated through explicit routing geometry.

The common obstacle is that global routing reports where demand lands for the current placement, but not how it would shift as cells move. To address this, we present GRADPlace, a geometry-based, routing-augmented differentiable placement framework. As shown in Fig.~\ref{fig:representative_vs_segment}, GRADPlace represents alternative rectilinear routes as probability-weighted directional segments. Their probabilities are estimated from global-routing feedback, while their vertices follow pin coordinates, so segment geometry, expected demand, and electrostatic congestion fields are recomputed at every placement step even when the probabilities stay fixed between routing refreshes. Integrating these fields over segments and propagating gradients through segment geometry directly links cell movement to demand redistribution.

The main contributions of this work are as follows:
\begin{itemize}[leftmargin=*]
\setlength{\emergencystretch}{1em}
\item A probability-weighted routing representation based on movable directional segments that preserves alternative rectilinear routing patterns and updates their geometry and expected directional demand with cell coordinates.
\item An electrostatic routing objective that integrates directional congestion fields over segment footprints and propagates the resulting gradients through segment geometry to connected cells.
\item An integrated placement flow combining segment-based routing gradients with cell inflation, routing-aware density, filler adjustment, and downstream refinement, achieving substantial DRV reductions across the ISPD 2015 benchmark suite with lower via count and comparable routed wirelength.
\end{itemize}

\section{Preliminaries}
\label{sec:PRELIMINARIES}

This section reviews continuous analytical placement and the discrete routing decisions modeled by GRADPlace.

\subsection{Analytical Placement with Electrostatic Density Modeling}
Modern analytical placers optimize cell locations by minimizing a continuous objective that combines wirelength and density regularization~\cite{lu2014eplace,cheng2019replace,lin2021dreamplace}. A standard formulation is
\[
\min_x \; W(x) + \lambda N_{\mathrm{density}}(x),
\]
where $x$ denotes the placement variables, $W(x)$ is the wirelength objective, and $N_{\mathrm{density}}(x)$ is a density penalty that discourages overlap and promotes a legalizable placement.

In electrostatics-based placement methods such as ePlace~\cite{lu2014eplace}, cells are modeled as movable charged objects distributed over the placement region. Let $\rho(z)$ denote the charge density at spatial coordinate $z$ over the placement domain $R$, and let $\phi(z)$ denote the corresponding potential. They satisfy the Poisson equation
\[
\nabla_z^2 \phi(z) = -\rho(z), \qquad z\in R,
\]
with suitable boundary conditions. The electric field is $\mathbf E(z)=-\nabla_z\phi(z)$, and the induced potential energy provides a smooth density regularizer for placement optimization. GRADPlace reuses this computational structure to construct a route-induced potential from moving routing-segment geometry.

\subsection{Routing as a Combinatorial Decision Problem}
\label{sec:routing_decisions}

Global routing includes discrete choices of paths and routing patterns. Pattern- and maze-routing-based routers expose this combinatorial structure through explicit pattern or path selection~\cite{xu2009fastroute4,pan2012fastroute,liu2013nctugr2,liu2020cugr}. To isolate these structural decisions, consider a finite candidate-path set $\mathcal K_s$ for each two-pin subnet $s$ and binary selection variables:
\[
\begin{aligned}
\min_{\{y_{s,k}\}}\quad
&\sum_s\sum_{k\in\mathcal K_s}c_{s,k}y_{s,k}\\
\text{s.t.}\quad
&\sum_{k\in\mathcal K_s}y_{s,k}=1,\qquad \forall s,\\
&y_{s,k}\in\{0,1\},\qquad \forall s,\ k\in\mathcal K_s.
\end{aligned}
\]
Here $y_{s,k}$ indicates whether candidate $k$ is selected for subnet $s$, and $c_{s,k}$ is its cost. This abstraction captures discrete path selection; a full routing formulation also couples these choices through shared capacity constraints.

Directly incorporating the full routing decision space into placement optimization is costly. GRADPlace therefore relaxes $y_{s,k}\in\{0,1\}$ into probabilities $p_{s,k}\in[0,1]$ over alternative rectilinear candidates on Steiner-connected subnets. Section~\ref{sec:Method} estimates these probabilities from routing feedback and develops their placement-dependent geometry.

\section{GRADPlace Method}
\label{sec:Method}

\begin{figure*}[t]
    \centering
    \includegraphics[width=\textwidth]{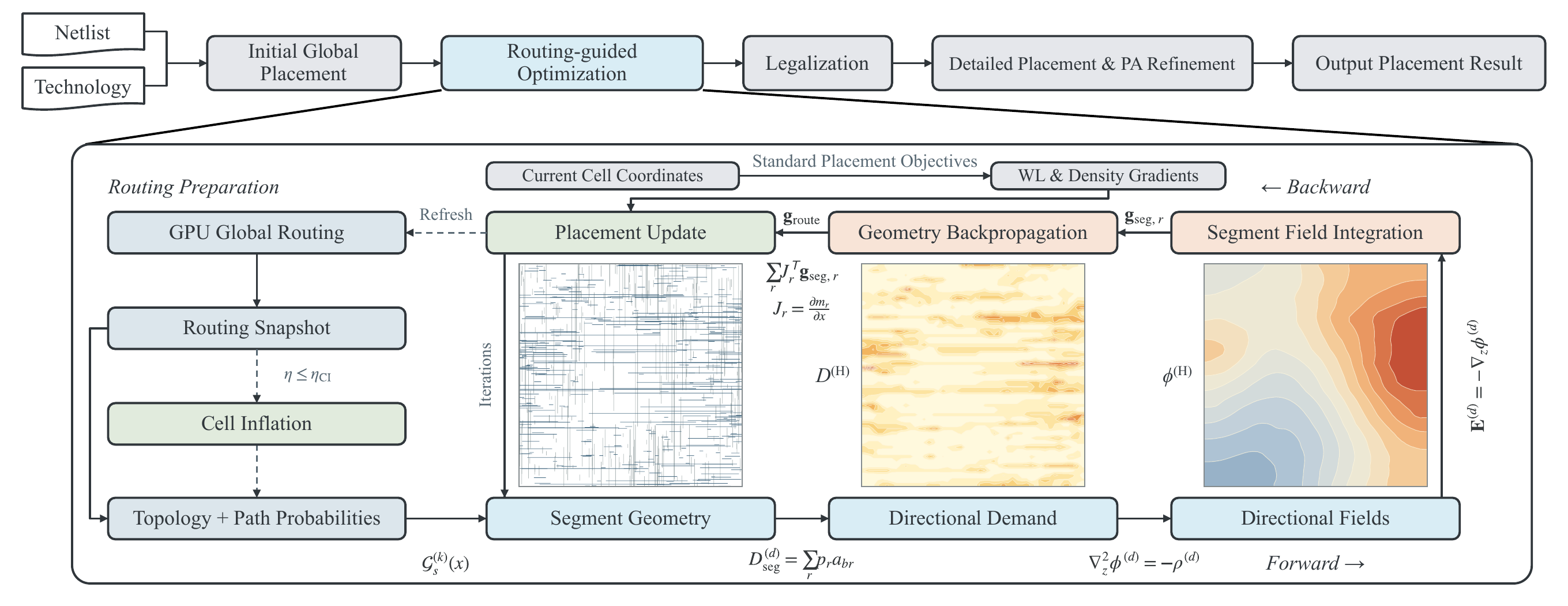}
    \caption{GRADPlace framework. Routing feedback determines candidate-path probabilities and triggers cell inflation when $\eta\leq\eta_{\mathrm{CI}}$, while movable directional segments define expected routing demand and electrostatic congestion fields. Field integration and geometric backpropagation supply cell gradients within the complete placement flow. Insets illustrate horizontal-resource segment geometry, expected demand $D^{(\mathrm H)}$, and potential $\phi^{(\mathrm H)}$.}
    \Description{The complete flow includes global placement, routing-guided optimization, legalization, and detailed placement with PA refinement. Routing snapshots supply path probabilities and optional threshold-triggered cell inflation. Current cell coordinates determine segments, directional demand, and congestion fields; segment field integrals propagate through segment geometry to cell coordinates.}
    \label{fig:framework_horizontal_overview}
\end{figure*}

GRADPlace models global congestion through a probability-weighted representation of movable directional segments. Fig.~\ref{fig:framework_horizontal_overview} summarizes the flow from candidate geometry to expected routing demand, capacity overflow, and an electrostatic routing objective. Routing feedback refreshes the path probabilities every $K_{\mathrm{GGR}}$ placement steps, an interval we call a routing epoch. Within an epoch, segment geometry and the resulting congestion model evolve with cell coordinates and provide gradients for analytical placement. We call the components of Sections~\ref{sec:routing_proxy}--\ref{sec:congestion_model} and the filler adjustment of Section~\ref{sec:filler_adjustment} the GRAD module; it includes the periodic GGR refreshes that update the path probabilities.

\subsection{Probabilistic Path Model and Movable Segment Geometry}
\label{sec:routing_proxy}

Inside the GPU-accelerated global router GGR~\cite{lin2022ggr}, a FLUTE-generated Steiner tree~\cite{chu2008flute} provides a compact connectivity skeleton, but it does not determine which routing resources the router actually uses. Because actual routing depends on grid capacity, blockages, and competing nets, a single deterministic Steiner geometry cannot capture the path uncertainty relevant to placement. GRADPlace therefore retains a compact candidate set $\mathcal K_s$ of placement-dependent H-first and V-first patterns for each Steiner-connected two-pin subnet $s$, as shown in Fig.~\ref{fig:gradplace_segments}. These patterns approximate plausible rectilinear realizations without enumerating all routing paths.

Each candidate carries a probability $p_{s,k}\geq 0$ with $\sum_{k\in\mathcal K_s}p_{s,k}=1$, and Section~\ref{sec:path_probability} estimates the matrix $\mathbf P=[p_{s,k}]_{s,k}$. Probability-weighted demand is thus an expectation over mutually exclusive candidates. Each candidate $k\in\mathcal K_s$ is decomposed into a set $\mathcal G_s^{(k)}(x)$ of placement-dependent horizontal and vertical directional segments. Because FLUTE Steiner points lie on the Hanan grid of the net's pins, every vertex $v$ inherits each coordinate from an associated pin:
\begin{equation}
X_v=\pi_{\iota_{\mathrm{x}}(v)}^{\mathrm{x}}(x),\qquad
Y_v=\pi_{\iota_{\mathrm{y}}(v)}^{\mathrm{y}}(x).
\label{eq:vertex_association}
\end{equation}
Here the pin indices $\iota_{\mathrm{x}}(v)$ and $\iota_{\mathrm{y}}(v)$ are fixed within a routing epoch, and $\pi_i^{\mathrm{x}}(x)$ and $\pi_i^{\mathrm{y}}(x)$ extract the coordinates of pin $i$ from $x$. A collinear subnet has a single candidate with $p_{s,1}=1$, and each segment $r\in\mathcal G_s^{(k)}(x)$ inherits $p_r=p_{s,k}$. Segment $r$ has a rectangular footprint $R_r(x)$ with midpoint $m_r(x)$ and length $\ell_r(x)$. Its width is one bin, $w_{\mathrm H}=\Delta_y$ for horizontal and $w_{\mathrm V}=\Delta_x$ for vertical segments.

\subsection{Path Probability Estimation from Routing Feedback}
\label{sec:path_probability}

GGR supplies routing feedback for both GRADPlace and Xplace-route, the routability-driven mode of Xplace~\cite{liu2024xplace}. At each routing epoch, GGR provides routed intervals and history congestion penalties to refresh $\mathbf P$.

For subnet $s$, let $\mathcal I_s$ contain the routed interval pieces of its net from the current GGR snapshot. Candidate support and topology cost are
\begin{equation}
\begin{aligned}
S_{s,k}&=\sum_{\gamma\in\mathcal I_s}\ell(\gamma)w_{s,k}(\gamma),\\
c_{s,k}^{\mathrm{topo}}
&=-\log\left(
\frac{S_{s,k}}{\sum_{k'\in\mathcal K_s}S_{s,k'}}+\epsilon_{\log}
\right),
\end{aligned}
\label{eq:topology_support}
\end{equation}
where $\ell(\gamma)$ is interval length, $\epsilon_{\log}>0$ guards against $\log 0$, and $w_{s,k}(\gamma)\in[0,1]$ is nonzero only when $\gamma$ lies on candidate $k$'s side of the subnet diagonal, decaying exponentially with the $L_1$ distance to the same-direction leg of $k$.

For subnets with $\sum_{k'\in\mathcal K_s}S_{s,k'}=0$, which have no topology support, a fallback cost scores $\mathcal G_s^{(k)}(x)$ against the directional congestion penalties $h_b^{(d)}$ recorded by GGR:
\begin{equation}
c_{s,k}^{\mathrm{cong}}
=\sum_{b,d}a_b^{(d)}\!\left(\mathcal G_s^{(k)}(x)\right)
h_b^{(d)},
\label{eq:path_congestion_cost}
\end{equation}
where $a_b^{(d)}(\mathcal G_s^{(k)}(x))$, defined in Section~\ref{sec:congestion_model}, is the candidate's direction-$d$ overlap area with bin $b$. We set $c_{s,k}=c_{s,k}^{\mathrm{topo}}$ when $\sum_{k'\in\mathcal K_s}S_{s,k'}>0$ and $c_{s,k}=c_{s,k}^{\mathrm{cong}}$ otherwise, and compute the path probabilities as
\begin{equation}
p_{s,k}
=\frac{\exp(-c_{s,k}/\tau_s)}
{\sum_{k'\in\mathcal K_s}\exp(-c_{s,k'}/\tau_s)},
\label{eq:path_probability}
\end{equation}
with temperature $\tau_s>0$ controlling distribution softness. Eq.~(\ref{eq:path_probability}) minimizes $\sum_{k}c_{s,k}p_{s,k}+\tau_s\sum_{k}p_{s,k}\log p_{s,k}$, an entropy-regularized relaxation of the selection problem in Section~\ref{sec:routing_decisions}.

\subsection{Electrostatic Congestion Modeling from Expected Demand}
\label{sec:congestion_model}

\subsubsection{Expected routing demand}
\label{sec:segment_demand}
Consider a demand grid with bin size $\Delta_x\times\Delta_y$ and bin area $A_{\mathrm{bin}}=\Delta_x\Delta_y$. Let $a_{br}=|R_r(x)\cap b|$ be the overlap area between segment $r$ and bin $b$, and $a_b^{(d)}(\mathcal G_s^{(k)}(x))=\sum_{r\in\mathcal G_s^{(k)},d(r)=d}a_{br}$. Expected directional demand under $\mathbf P$ is
\begin{equation}
D_{\mathrm{seg},b}^{(d)}(x;\mathbf P)
=\sum_s\sum_{k\in\mathcal K_s}p_{s,k}
a_b^{(d)}\!\left(\mathcal G_s^{(k)}(x)\right).
\label{eq:segment_demand}
\end{equation}
Since segments are one bin wide, a horizontal segment that overlaps bin $b$ over length $\ell_{br}$ contributes $a_{br}/A_{\mathrm{bin}}=\ell_{br}/\Delta_x\leq 1$ track to that bin, and vertical segments are analogous.

\subsubsection{Capacity overflow and electrostatic fields}
\label{sec:route_field}
GGR supplies directional capacity $C_b^{(d)}$ and fixed occupancy $B_b^{(d)}$ from fixed routing usage and blockages. Both are in track units and resampled onto the demand grid, then combined with expected segment demand:
\begin{equation}
\begin{aligned}
o_b^{(d)}(x)&=B_b^{(d)}
+\frac{D_{\mathrm{seg},b}^{(d)}(x;\mathbf P)}{A_{\mathrm{bin}}},\\
q_b^{(d)}(x)&=
\left[
\frac{o_b^{(d)}(x)-C_b^{(d)}}
{\max(C_b^{(d)},\epsilon_C)}
\right]_+,
\end{aligned}
\label{eq:capacity_overflow}
\end{equation}
with $\epsilon_C>0$ guarding zero-capacity bins. Within a routing epoch, moving segments continually update $D_{\mathrm{seg},b}^{(d)}$ and hence the normalized overflow source $q_b^{(d)}$.

Following electrostatic placement~\cite{lu2014eplace,zhu2018poisson}, we subtract the spatial mean from each directional congestion source and solve the Poisson equation separately for horizontal and vertical resources:
\begin{equation}
\begin{aligned}
\rho^{(d)}&=(q^{(d)}-\overline{q^{(d)}})/A_{\mathrm{bin}},\\
\nabla_z^2\phi^{(d)}(z)&=-\rho^{(d)}(z),\qquad
\mathbf E^{(d)}(z)=-\nabla_z\phi^{(d)}(z).
\end{aligned}
\label{eq:route_poisson}
\end{equation}
Dividing by $A_{\mathrm{bin}}$ gives a charge density in the units of the placement Poisson solve, and the DCT solver suppresses the zero-frequency mode. The resulting fields push demand from overloaded regions toward available resources.

\subsubsection{Gradients through segment geometry}
\label{sec:segment_backward}
At each placement step, segment geometry and directional potentials are recomputed from the current placement. Together with the epoch-level probabilities, they define the routing objective
\begin{equation}
N_{\mathrm{route}}(x)
=\sum_r p_r
\int_{R_r(x)}
\phi^{(d(r))}(z)\,dz,
\label{eq:route_objective}
\end{equation}
where $d(r)\in\{\mathrm H,\mathrm V\}$ denotes the direction of segment $r$.

When differentiating Eq.~(\ref{eq:route_objective}), each segment is treated as a rigid charge within the current step: $\phi^{(d)}$, $p_r$, and $\ell_r$ are held fixed and only the midpoint $m_r(x)$ moves, mirroring the constant cell charges of electrostatic placement~\cite{lu2014eplace}. Because $\phi^{(d)}$ is defined only up to an additive constant, differentiating $\ell_r$ would add a convention-dependent multiple of the wirelength gradient; fixing $\ell_r$ avoids this and leaves length changes to $W(x)$. The resulting frozen-field gradient is
\begin{equation}
\nabla_xN_{\mathrm{route}}\approx\sum_r J_r^{\top}\mathbf g_{\mathrm{seg},r},\quad
\mathbf g_{\mathrm{seg},r}=-p_r\sum_b a_{br}\mathbf E_b^{(d(r))},
\label{eq:segment_gradient}
\end{equation}
with $J_r=\partial m_r/\partial x$ given by Eq.~(\ref{eq:vertex_association}). The along-segment force is thus split equally between the two end vertices, which removes the length component of the gradient and remains continuous when endpoint pins cross.

\subsection{Integration with Placement Optimization}
\label{sec:route_optimization}

GRADPlace combines the segment-based routing gradient with the standard analytical-placement terms. At every placement step, the update direction is formed from
\begin{equation}
\mathbf g
=\nabla_x W(x)
+\lambda_1\nabla_xN_{\mathrm{density}}(x)
+\lambda_2\nabla_xN_{\mathrm{route}}(x).
\label{eq:placement_gradient}
\end{equation}
Here $W$ and $N_{\mathrm{density}}$ are the wirelength and density objectives; $\lambda_1$ and $\lambda_2$ balance density and routability.

\subsubsection{Routing-aware density terms}
Following Xplace~\cite{liu2024xplace}, our placement density model also accounts for routing-related occupancy that is not represented by movable-cell area alone. For each bin $b$ with area $A_{\mathrm{bin}}$, the I/O-pin and M2-rail densities are
\begin{equation}
\begin{aligned}
D_{\mathrm{iopin},b}
&=\frac{\sum_{u\in\mathcal V_{\mathrm{iopin}}} \lvert A_u\cap b\rvert}{A_{\mathrm{bin}}},\\
D_{\mathrm{m2rail},b}
&=\frac{\sum_{u\in\mathcal V_{\mathrm{m2rail}}} \lvert A_u\cap b\rvert}{A_{\mathrm{bin}}}.
\end{aligned}
\label{eq:routing_density_terms}
\end{equation}
Let $D_b$ be the ordinary placement density of bin $b$, given by the fraction of bin area covered by placed instances. Following Xplace~\cite{liu2024xplace}, we augment it as
\begin{equation}
\widetilde D_b
=D_b+D_{\mathrm{m2rail},b}+3D_{\mathrm{iopin},b},
\label{eq:routing_density}
\end{equation}
and use this routing-aware density map $\widetilde D_b$ in $N_{\mathrm{density}}$.

\subsubsection{Auxiliary filler adjustment}
\label{sec:filler_adjustment}
Routing-driven cell movement can leave underfilled regions, and the density term then pulls cells back. We therefore move nearby fillers into the vacated space using the aggregate directional field. Fillers have no pins or routing segments, so this adjustment adds no routing demand. The field force on filler $f$ is
\begin{equation}
\mathbf r_f=\int_{S_f}\mathbf E_\Sigma(z)\,dz
\;\approx\;\sum_b |S_f\cap b|\,\mathbf E_{\Sigma,b}.
\label{eq:filler_field_term}
\end{equation}
Here $S_f$ is the filler footprint and $\mathbf E_\Sigma=\mathbf E^{(\mathrm H)}+\mathbf E^{(\mathrm V)}$. With a positive diagonal preconditioner $M_f$, the filler descent direction is
\begin{equation}
\widehat{\mathbf g}_f=M_f^{-1}\!\left(\lambda_1\nabla_{x_f}N_{\mathrm{density}}
+\lambda_f\mathbf r_f\right).
\label{eq:filler_auxiliary_update}
\end{equation}
The term $\lambda_f\mathbf r_f$ drives fillers against $\mathbf E_\Sigma$, toward the congested, high-potential bins that movable cells have vacated.

\subsubsection{Cell inflation and placement refinement}
Cell inflation uses routing feedback to adjust effective cell areas, encouraging local spreading through $N_{\mathrm{density}}$, while the segment-based objective moves connected cells so that their candidate routes leave congested bins. Algorithm~\ref{alg:gradplace} summarizes the flow, where $\eta^{(i)}$ denotes placement overflow at step $i$. Routing-augmented optimization starts once initial global placement reaches $\eta^{(0)}\leq\eta_{\mathrm{route}}$, and cell inflation is triggered whenever $\eta^{(i)}\leq\eta_{\mathrm{CI}}$, up to $N_{\mathrm{CI}}^{\max}$ rounds. Tool-native legalization, ABCDPlace detailed placement~\cite{lin2020abcdplace}, and the pin-access (PA) refinement shared with Xplace-route then produce the final legal placement.

\begin{algorithm}[t]
\caption{GRADPlace Routing-Augmented Optimization}
\label{alg:gradplace}
\small
\begin{algorithmic}[1]
\State \textbf{Input:} $x^{(0)}$, $M$, $N_{\mathrm{CI}}^{\max}$, $\eta_{\mathrm{stop}}$, $\eta_{\mathrm{route}}$, $\eta_{\mathrm{CI}}$, and $K_{\mathrm{GGR}}$
\State Run initial global placement until overflow $\eta^{(0)} \leq \eta_{\mathrm{route}}$
\State Initialize Steiner topology and path probability $\mathbf P$ by GGR
\State $i \leftarrow 0$, $n_{\mathrm{CI}} \leftarrow 0$
\While{$i<M$ and $\eta^{(i)}>\eta_{\mathrm{stop}}$}
    \If{$i>0$ and $i \bmod K_{\mathrm{GGR}} = 0$}
        \State Rebuild Steiner topology and $\mathbf P$ by GGR
    \EndIf
    \State Compute expected routing demand $D_{\mathrm{seg}}$ from $x^{(i)}$ and $\mathbf P$
    \State Compute routing gradient $\nabla_xN_{\mathrm{route}}$ via Eq.~(\ref{eq:segment_gradient})
    \State Update $x^{(i+1)}$ by one Nesterov step using $\mathbf g$ and $\widehat{\mathbf g}_f$
    \If{$\eta^{(i)}\leq\eta_{\mathrm{CI}}$ and $n_{\mathrm{CI}} < N_{\mathrm{CI}}^{\max}$}
        \State Invoke GGR, apply cell inflation, and set $n_{\mathrm{CI}} \leftarrow n_{\mathrm{CI}}+1$
    \EndIf
    \State $i \leftarrow i+1$
\EndWhile
\State \Return $x^{(i)}$
\end{algorithmic}
\end{algorithm}

\section{Results}
\label{sec:results}

\begin{table*}[t]
\centering
\caption{ISPD 2015 comparison routed by Innovus~\cite{cadence_innovus}. Mean ratio denotes the arithmetic mean of per-design ratios to GRADPlace.}
\label{tab:gradplace-pci-comparison}
\setlength{\tabcolsep}{2.2pt}
\renewcommand{\arraystretch}{0.78}
\scriptsize
\resizebox{\textwidth}{!}{%
\begin{tabular}{p{2.4cm}r*{15}{r}}
\toprule
Case & \#Cells (K) & \multicolumn{5}{c}{DCGP$^\ddagger$} & \multicolumn{5}{c}{Xplace-route} & \multicolumn{5}{c}{GRADPlace} \\
\cmidrule(lr){3-7}\cmidrule(lr){8-12}\cmidrule(lr){13-17}
 &  & DR WL ($\mu$m) & \#Vias & \#DRVs & \textcolor{gray}{Place (s)} & \textcolor{gray}{Route (s)} & DR WL ($\mu$m) & \#Vias & \#DRVs & Place (s) & Route (s) & DR WL ($\mu$m) & \#Vias & \#DRVs & Place (s) & Route (s) \\
\midrule
mgc\_des\_perf\_1 & 112.6 & \textbf{1,429,056} & 566,581 & 10,652 & \textcolor{gray}{30} & \textcolor{gray}{568} & 1,458,591 & 577,516 & 11,809 & \textbf{18} & \textbf{905} & 1,470,315 & \textbf{553,937} & \textbf{2,202} & 153 & 952 \\
mgc\_des\_perf\_a\smash{\rlap{\textsuperscript{\(\dagger\)}}} & 108.3 & 2,360,122 & 569,940 & 10,094 & \textcolor{gray}{20} & \textcolor{gray}{3,430} & 2,440,095 & 573,370 & 45,202 & \textbf{10} & \textbf{272} & \textbf{2,299,353} & \textbf{564,472} & \textbf{5,027} & 117 & 2,340 \\
mgc\_des\_perf\_b\smash{\rlap{\textsuperscript{\(\dagger\)}}} & 112.6 & 1,783,673 & 541,130 & 849 & \textcolor{gray}{21} & \textcolor{gray}{89} & \textbf{1,768,510} & \textbf{533,368} & 2,108 & \textbf{5} & 210 & 1,868,361 & 537,633 & \textbf{803} & 49 & \textbf{121} \\
mgc\_edit\_dist\_a\smash{\rlap{\textsuperscript{\(\dagger\)}}} & 127.4 & 5,737,250 & 1,034,387 & 359,904 & \textcolor{gray}{19} & \textcolor{gray}{801} & 5,725,110 & \textbf{997,993} & 423,181 & \textbf{13} & 983 & \textbf{5,699,748} & 1,015,150 & \textbf{314,636} & 210 & \textbf{852} \\
mgc\_fft\_1 & 32.3 & 513,993 & 185,309 & 3,661 & \textcolor{gray}{26} & \textcolor{gray}{299} & 510,202 & 181,490 & 3,533 & \textbf{5} & 341 & \textbf{500,892} & \textbf{173,742} & \textbf{934} & 94 & \textbf{152} \\
mgc\_fft\_2 & 32.3 & 612,371 & 190,620 & 1,070 & \textcolor{gray}{14} & \textcolor{gray}{328} & 594,589 & 187,848 & 2,677 & \textbf{2} & 521 & \textbf{552,388} & \textbf{170,991} & \textbf{719} & 54 & \textbf{101} \\
mgc\_fft\_a & 30.6 & 1,133,734 & 192,225 & 790 & \textcolor{gray}{9} & \textcolor{gray}{273} & 1,115,228 & 189,297 & 1,512 & \textbf{2} & 392 & \textbf{1,085,301} & \textbf{182,087} & \textbf{683} & 67 & \textbf{230} \\
mgc\_fft\_b & 30.6 & 1,310,187 & 216,838 & 14,710 & \textcolor{gray}{9} & \textcolor{gray}{263} & 1,303,070 & 212,225 & 15,796 & \textbf{6} & 424 & \textbf{1,258,116} & \textbf{202,997} & \textbf{7,862} & 111 & \textbf{285} \\
mgc\_matrix\_mult\_1 & 155.3 & 2,638,437 & 820,336 & 10,717 & \textcolor{gray}{25} & \textcolor{gray}{2,180} & 2,616,882 & 809,003 & 13,450 & \textbf{13} & 1,317 & \textbf{2,591,046} & \textbf{784,612} & \textbf{3,459} & 209 & \textbf{1,067} \\
mgc\_matrix\_mult\_2 & 155.3 & 2,676,434 & 853,057 & 10,651 & \textcolor{gray}{26} & \textcolor{gray}{2,975} & 2,672,784 & 848,638 & 14,364 & \textbf{13} & 3,671 & \textbf{2,603,537} & \textbf{784,319} & \textbf{2,688} & 214 & \textbf{588} \\
mgc\_matrix\_mult\_a & 149.7 & 3,938,218 & 849,468 & 9,304 & \textcolor{gray}{18} & \textcolor{gray}{1,704} & \textbf{3,919,935} & 835,604 & 7,212 & \textbf{7} & 1,952 & 3,921,211 & \textbf{802,687} & \textbf{2,018} & 201 & \textbf{316} \\
mgc\_matrix\_mult\_b\smash{\rlap{\textsuperscript{\(\dagger\)}}} & 146.4 & 3,680,673 & 789,855 & 37,698 & \textcolor{gray}{18} & \textcolor{gray}{323} & \textbf{3,654,794} & \textbf{772,167} & 44,507 & \textbf{6} & \textbf{501} & 3,721,617 & 776,528 & \textbf{8,658} & 169 & 4,599 \\
mgc\_matrix\_mult\_c\smash{\rlap{\textsuperscript{\(\dagger\)}}} & 146.4 & 3,679,849 & 795,182 & 8,907 & \textcolor{gray}{19} & \textcolor{gray}{686} & 3,675,958 & 780,732 & 7,597 & \textbf{6} & 1,205 & \textbf{3,644,203} & \textbf{744,917} & \textbf{2,222} & 270 & \textbf{280} \\
mgc\_pci\_bridge32\_a\smash{\rlap{\textsuperscript{\(\dagger\)}}} & 29.5 & 649,379 & 146,588 & 3,660 & \textcolor{gray}{9} & \textcolor{gray}{1,035} & \textbf{620,149} & 140,304 & 1,979 & \textbf{2} & 669 & 644,054 & \textbf{139,445} & \textbf{1,937} & 17 & \textbf{632} \\
mgc\_pci\_bridge32\_b\smash{\rlap{\textsuperscript{\(\dagger\)}}} & 28.9 & 998,056 & 148,052 & 198 & \textcolor{gray}{9} & \textcolor{gray}{38} & 992,290 & 142,786 & 359 & \textbf{3} & \textbf{96} & \textbf{967,068} & \textbf{138,222} & \textbf{194} & 42 & 106 \\
mgc\_superblue11\_a\smash{\rlap{\textsuperscript{\(\dagger\)}}} & 925.6 & \textbf{40,209,334} & 5,652,221 & \textbf{669} & \textcolor{gray}{134} & \textcolor{gray}{1,670} & 40,253,744 & 5,548,309 & 1,277 & \textbf{41} & 2,311 & 41,259,825 & \textbf{5,445,328} & 947 & 663 & \textbf{1,773} \\
mgc\_superblue12 & 1286.9 & 49,383,559 & 12,779,668 & 15,113 & \textcolor{gray}{260} & \textcolor{gray}{6,796} & \textbf{43,581,188} & 10,404,590 & 23,568 & \textbf{184} & 7,108 & 44,554,200 & \textbf{9,827,110} & \textbf{3,889} & 2,895 & \textbf{3,985} \\
mgc\_superblue14 & 612.2 & 28,097,546 & 4,289,310 & \textbf{310} & \textcolor{gray}{147} & \textcolor{gray}{1,511} & 28,074,428 & 4,281,616 & 423 & \textbf{31} & 2,498 & \textbf{27,718,344} & \textbf{3,850,515} & 336 & 925 & \textbf{1,284} \\
mgc\_superblue16\_a\smash{\rlap{\textsuperscript{\(\dagger\)}}} & 680.5 & \textbf{31,267,452} & 4,653,129 & 3,387 & \textcolor{gray}{157} & \textcolor{gray}{4,004} & 31,597,960 & \textbf{4,581,054} & 4,970 & \textbf{36} & 5,691 & 31,765,742 & 4,992,028 & \textbf{2,718} & 328 & \textbf{3,607} \\
mgc\_superblue19 & 506.1 & 21,126,469 & 3,647,126 & 8,280 & \textcolor{gray}{109} & \textcolor{gray}{1,781} & \textbf{20,733,972} & \textbf{3,508,860} & 5,505 & \textbf{37} & 2,295 & 21,095,507 & 3,563,225 & \textbf{2,114} & 483 & \textbf{1,648} \\
\midrule
Mean ratio & -- & 1.018 & 1.058 & 2.555 & \textcolor{gray}{0.202} & \textcolor{gray}{1.704} & 1.006 & 1.029 & 3.370 & \textbf{0.065} & 2.129 & \textbf{1.000} & \textbf{1.000} & \textbf{1.000} & 1.000 & \textbf{1.000} \\
\bottomrule
\end{tabular}%
}
\par\vspace{3pt}
{\footnotesize\raggedright

$^\ddagger$~Xplace-route and GRADPlace use the evaluation script from Xplace~\cite{liu2024xplace}. DCGP quality metrics are quoted from its published report using the same script; Table~\ref{tab:xplace-script-check} supports the consistency of this evaluation. DCGP runtime, shown in gray, was measured on different hardware and is excluded from bold runtime comparison.

$^\dagger$~marks benchmark cases with fence-region constraints removed. \par}
\end{table*}

\subsection{Experimental Setup}
We evaluate GRADPlace on all 20 designs of the ISPD 2015 benchmark suite~\cite{bustany2015ispd} using detailed-routing wirelength (DR WL), via count, design-rule violations after detailed routing (\#DRVs), and placement and routing runtimes. Xplace-route and GRADPlace use tool-native legalization, the same ABCDPlace detailed placement (DP)~\cite{lin2020abcdplace}, and PA refinement under our Innovus~\cite{cadence_innovus} flow. DCGP quality metrics are quoted from its published report, which uses the same script and router; as Table~\ref{tab:xplace-script-check} shows, its Xplace-route results agree with the original Xplace report and our reproduction, so these metrics are comparable with ours. DCGP runtime was measured on different hardware and is not used for efficiency comparison. Placement time covers global placement with its in-loop GGR calls, legalization, ABCDPlace, and PA refinement; route time covers Innovus global and detailed routing. Following Xplace~\cite{liu2024xplace} and DCGP~\cite{li2025dcgp}, we remove fence-region constraints from the benchmarks marked $^\dagger$ for all methods. GRADPlace is built on DREAMPlace. Our DREAMPlace cell-inflation baseline, DREAMPlace+GGR, applies GGR-driven cell inflation without the GRAD module, calling GGR only at inflation triggers, and is used in the ablation and case studies.

All runs use deterministic execution with random seed 3000 on AMD EPYC 9654 CPUs and one NVIDIA RTX 6000 Ada GPU. Table~\ref{tab:gradplace-pci-comparison} reports the cross-method comparison, and Table~\ref{tab:hyperparameters} lists the fixed GRADPlace hyperparameters used across all benchmarks.

\begin{table}[t]
\caption{GRADPlace settings used for all experiments.}
\label{tab:hyperparameters}
\centering
\small
\renewcommand{\arraystretch}{1.05}
\begin{tabular}{@{}p{0.32\columnwidth} p{0.60\columnwidth}@{}}
\toprule
Hyperparameter & Setting \\
\midrule
$M$ & 3000 placement steps \\
$N_{\mathrm{CI}}^{\max}$ & 5 cell-inflation rounds \\
$\eta_{\mathrm{route}}$ & $0.3$ placement overflow \\
$\eta_{\mathrm{CI}}$ & $0.15$ placement overflow \\
$\eta_{\mathrm{stop}}$ & $0.07$ placement overflow \\
$K_{\mathrm{GGR}}$ & 10 placement steps \\
$\tau_s$ & $0.2\cdot\operatorname{mean}_{k\in\{\mathrm H,\mathrm V\}} c_{s,k}$ \\
$\lambda_1$ & DREAMPlace density-weight update \\
$\lambda_2$ & $0.2\cdot\dfrac{\lVert\nabla_x(W(x)+\lambda_1N_{\mathrm{density}}(x))\rVert}{\lVert\nabla_xN_{\mathrm{route}}(x)\rVert}$ \\
$\lambda_f$ & $0.2\cdot\dfrac{\lVert\nabla_{x_f}N_{\mathrm{density}}(x)\rVert}{\lVert\mathbf r_f\rVert}$ \\
\bottomrule
\end{tabular}
\end{table}

\begin{table}[t]
\centering
\caption{Ablation on all 20 ISPD 2015 designs, reported as mean ratios to the full flow.}
\label{tab:gradplace-compact-ablation}
\setlength{\tabcolsep}{5pt}
\renewcommand{\arraystretch}{1.08}
\small
\resizebox{0.75\columnwidth}{!}{%
\begin{tabular}{ccc|ccc}
\toprule
\multicolumn{3}{c|}{Methods} & \multicolumn{3}{c}{Mean ratio to full flow} \\
\cmidrule(lr){1-3}\cmidrule(lr){4-6}
CI & GRAD & DP+PA & DR WL & \#Vias & \#DRVs \\
\midrule
-- & -- & -- & 1.004 & 1.056 & 35.473 \\
\ding{51} & -- & -- & 1.007 & 1.053 & 5.131 \\
\ding{51} & \ding{51} & -- & 1.015 & 1.052 & 1.117 \\
\ding{51} & -- & \ding{51} & 1.001 & 1.003 & 3.171 \\
\ding{51} & \ding{51} & \ding{51} & \textbf{1.000} & \textbf{1.000} & \textbf{1.000} \\
\bottomrule
\end{tabular}%
}
\end{table}

\subsection{Ablation Study}
Table~\ref{tab:gradplace-compact-ablation} evaluates the complete GRAD module across all 20 designs, both after legalization only and after the common DP+PA stages. Here, CI denotes GGR-driven cell inflation, and row 1 is the original DREAMPlace, which optimizes only wirelength and density without GGR feedback. Adding GRAD reduces the mean DRV ratio from 5.131 to 1.117 before DP+PA and from 3.171 to 1.000 after DP+PA, while routed wirelength and via count stay close to those of the full flow. The in-house CI+DP+PA baseline reaches a mean DRV ratio of 3.171, close to the 3.370 of Xplace-route in Table~\ref{tab:gradplace-pci-comparison}, so the DRV gains are attributable to GRAD and not to the underlying placement engine.

\subsection{Cross-method Comparison}
In Table~\ref{tab:gradplace-pci-comparison}, the mean per-design DRV ratios of Xplace-route and DCGP to GRADPlace are 3.370 and 2.555, and the corresponding geometric means are 2.854 and 2.124. GRADPlace has fewer DRVs than Xplace-route on all 20 designs and than DCGP on 18 of them. Xplace-route and DCGP also use 2.9\% and 5.8\% more vias on average, with comparable routed wirelength.

\begin{figure}[t]
\centering
\includegraphics[width=\columnwidth]{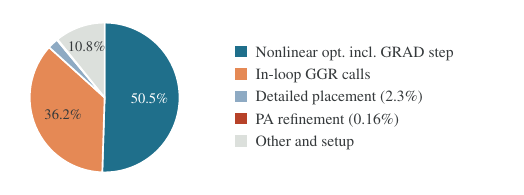}
\caption{Total placement runtime breakdown across the ISPD 2015 suite.}
\label{fig:placement-runtime-breakdown}
\end{figure}

Per-step field solves and in-loop GGR refreshes increase placement time over Xplace-route, whereas cleaner placements shorten detailed routing. GRADPlace lowers total place-and-route time on 14 of 20 designs, with a mean Xplace-route-to-GRADPlace ratio of 1.475. This comparison is conservative because Innovus can abort detailed routing early on unroutable placements~\cite{liu2024xplace}, which shortens the baseline route time on designs such as \textit{mgc\_des\_perf\_a} and \textit{mgc\_matrix\_mult\_b}. Fig.~\ref{fig:placement-runtime-breakdown} breaks down total placement time across all 20 designs: nonlinear optimization, including the GRAD step, takes 50.5\%, and in-loop GGR calls take 36.2\%.

\begin{table}[t]
\caption{Xplace-route mean ratios to GRADPlace.}
\label{tab:xplace-script-check}
\centering
\small
\renewcommand{\arraystretch}{1.05}
\begin{tabular}{@{}l rrr@{}}
\toprule
Xplace-route source & DR WL & \#Vias & \#DRVs \\
\midrule
Original Xplace report~\cite{liu2024xplace} & 1.013 & 1.046 & 3.220 \\
Reported in DCGP~\cite{li2025dcgp} & 1.018 & 1.049 & 3.357 \\
Our reproduction & 1.006 & 1.029 & 3.370 \\
\bottomrule
\end{tabular}
\end{table}

\begin{figure}[t]
\centering
\includegraphics[width=\columnwidth]{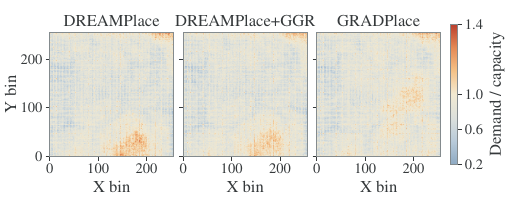}
\caption{GGR union congestion maps on \textit{mgc\_matrix\_mult\_1} for DREAMPlace, DREAMPlace+GGR, and GRADPlace. Orange and red bins exceed routing capacity.}
\Description{Three side-by-side congestion maps for the mgc matrix mult 1 design, comparing DREAMPlace, DREAMPlace+GGR, and GRADPlace on a shared congestion-ratio color scale.}
\label{fig:matrix-mult-congestion-comparison}
\end{figure}

\begin{figure}[!t]
\centering
\includegraphics[width=\columnwidth]{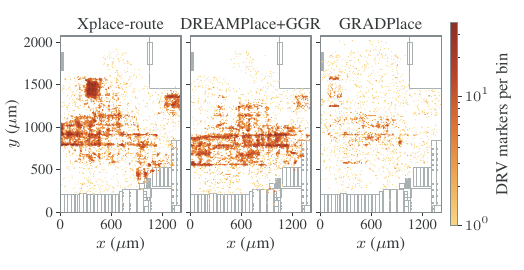}
\caption{Spatial distribution of DRV markers on \textit{mgc\_superblue12}. Gray outlines mark macros.}
\Description{Three placement panels compare the spatial distribution of DRV markers for Xplace-route, DREAMPlace+GGR, and GRADPlace on the mgc\_superblue12 design.}
\label{fig:superblue12-drc-spatial}
\end{figure}

\begin{figure}[!tb]
\centering
\includegraphics[width=\columnwidth]{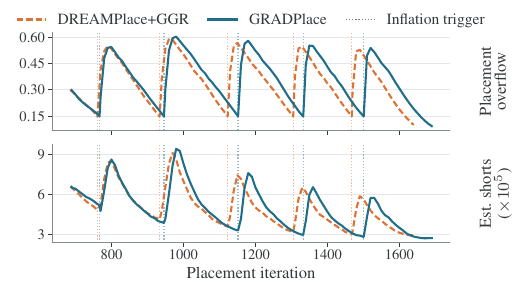}
\caption{Placement overflow and GGR estimated shorts during placement on \textit{mgc\_superblue12} for DREAMPlace+GGR and GRADPlace. Dotted lines mark inflation triggers.}
\Description{Two panels show placement overflow and GGR estimated shorts versus placement iteration. Dotted lines mark inflation triggers. GRADPlace has higher estimated shorts at the first trigger and lower values at triggers two through five.}
\label{fig:inflation-trigger-comparison}
\end{figure}

\subsection{Case Studies}
\label{sec:congestion-evolution}
Fig.~\ref{fig:matrix-mult-congestion-comparison} shows that GRADPlace flattens the concentrated bottom congestion hotspot that persists under DREAMPlace and DREAMPlace+GGR by redistributing demand into surrounding capacity. On \textit{mgc\_superblue12}, Fig.~\ref{fig:superblue12-drc-spatial} shows far sparser violation clusters under GRADPlace, which has 3,889 DRVs against 23,568 for Xplace-route and 19,517 for DREAMPlace+GGR.

Fig.~\ref{fig:inflation-trigger-comparison} compares GRADPlace with the cell-inflation baseline DREAMPlace+GGR on \textit{mgc\_superblue12}. Inflation triggers occur near $\eta=0.15$. At each trigger, inflating cell areas increases local overlap and resets the density weight, producing sharp temporary rises before cells re-spread.

At the first inflation trigger, GRADPlace has more estimated shorts because routing guidance has only just started. At triggers 2--5, its estimated shorts are 6.5--7.6\% lower than those of DREAMPlace+GGR, and they remain 7.1\% lower at the final GGR refresh.

\section{Conclusion}
\label{sec:Conclusion}

{\setlength{\emergencystretch}{1em}
GRADPlace introduces a probability-weighted routing representation based on movable directional segments for routability-driven analytical placement. At each placement step, current segment geometry determines expected directional demand and directional congestion fields, and an electrostatic routing objective propagates gradients through segment geometry to connected cells. Across the ISPD 2015 benchmark suite, GRADPlace substantially reduces design-rule violations and via counts relative to reproduced Xplace-route and published DCGP results while maintaining comparable routed wirelength. Current limitations include H/V-only candidate patterns, fixed weighting schedules, single-seed runs, and evaluation on one benchmark family and routing flow; future work will explore multi-bend candidates, adaptive schedules, and timing-coupled objectives.\par}

\bibliographystyle{IEEEtran}
\bibliography{main}

\end{document}